\documentclass[12pt,preprint]{aastex}

\newcommand{\myemail}{swolf@mpia.de}

\shorttitle{Giant planets in protoplanetary disks}
\shortauthors{Wolf \& D'Angelo}

\begin{document}


\title{On the Observability of Giant Protoplanets in Circumstellar Disks}


\author{Sebastian Wolf}
\affil{Max Planck Institute for Astronomy, K\"onigstuhl 17, 69117 Heidelberg,
  Germany}
\affil{California Institute of Technology, 1201 E California Blvd,\\ Mail code 105-24, Pasadena, CA 91125, USA}
\email{\myemail}

\and

\author{Gennaro D'Angelo\altaffilmark{1}}
\affil{School of Physics, University of Exeter,\\ Stocker Road, Exeter EX4 4QL, UK}
\email{gennaro@astro.ex.ac.uk}


\altaffiltext{1}{UKAFF Fellow.}


\begin{abstract}
We investigate the possibility to detect giant planets that are still embedded in young circumstellar disks.
Based on models with different stellar, planetary, and disk masses, and different radial positions
of the planet we analyze the resulting submillimeter appearance of these systems. 
We find that the influence of the planet on the spectral energy distribution could not be distinguished 
from that of other disk parameters. However, dust reemission {\em images} of the disks show that the hot region 
in the proximity of a young planet, along with the gap, could indeed be detected
and mapped with the Atacama Large Millimeter Array in the case of nearby circumstellar disks (d$<$100\,pc)
in approximate face-on orientation.
\end{abstract}




$ $ \newpage
\section{Introduction}\label{intro}

Planets are expected to form in circumstellar disks, which are considered as the natural outcome
of the protostellar evolution, at least in the case of low and medium mass stars
(e.g., Adams, Lada, \& Shu~1987; Lissauer~1993). While a detailed picture of the evolution of 
the circumstellar environment, in particular of circumstellar disks, has been developed already,
the planet formation process is mostly still under discussion.
There exist two main different scenarios for planet formation.
The first is characterized by three major phases:
(1)~dust grain growth from submicron-sized particles to centimeter /
decimeter-sized bodies via coagulation is followed by
(2)~an agglomeration process that leads to the formation of (sub)kilometer-sized planetesimals, which
(3)~form terrestrial (rocky) and Uranian (icy) planets or the cores of
Jovian (gaseous) planets by further accretion of solid material
(e.g., Pollack et al. 1996, Weidenschilling 1997).
Alternatively, planetesimals may form via the gravitational
instability of solids that have settled to the midplane 
of a circumstellar disk (Goldreich \& Ward~1973, Youdin \& Shu~2002).
Adequate constraints from observations are required in order to
either verify or rule out 
existing hypotheses about these planet formation scenarios.

During recent years, numerical simulations studying planet-disk interactions have shown
that planets may cause characteristic large-scale signatures in the disk density distributions.
The most important of these signatures are gaps and spiral density waves in young circumstellar
disks 
(e.g.,
Bryden et al.~1999;
Kley~1999;
Lubow, Seibert, \& Artymowicz~1999;
Kley, D'Angelo, \& Henning~2001;
Bate et al.~2003;
Winters, Balbus, \& Hawley~2003;
Nelson \& Papaloizou~2003) 
and resonance structures in evolved systems, so-called Debris disks 
(e.g.,
Liou \& Zook~1999,
Ozernoy et al.~2000,
Moro-Mart\'{\i}n \& Malhotra~2002).
The importance of investigating these signatures lies in the 
possibility that they can be used to search for embedded (i.e., young) planets. 
Therefore, disk features can provide constraints on the processes and timescales
of planet formation.
Indeed, several debris disks around nearby main-sequence stars show
structures and asymmetries that are considered to result from planetary perturbations 
(Holland et al.~1998, 2003;
Schneider et al.~1999;
Koerner, Sargent, \& Ostroff~2001).

In our present study we consider young disks with a density structure that is dominated
by gas dynamics. While it was shown before, that the planned Atacama Large Millimeter Array (ALMA)
will be able to map the gap caused by a massive planet in such disks (Wolf et al.~2002),
we are now using more detailed simulations in order to investigate whether the planet itself and/or
its surrounding environment could be detected.
The detection of a gap would already represent a strong indication of
the existence of a planet,
thus giving information such as planetary mass, viscosity, and
pressure scale-height of the disk.
The detection / non-detection of warm dust close
to the planet, however, would additionally provide valuable indications on the temperature and luminosity of the planet
and on the density structure of the surrounding medium.

We base our investigation on hydrodynamical simulations of circumstellar disks with an embedded planet
and subsequent radiative transfer simulations, with the aim of
deriving observable quantities of these systems.
Technical details on the performed simulations together with an
explanation and justification of the chosen 
model setup are given in \S~\ref{basics}.
The results are discussed in \S~\ref{results} and our conclusions are
presented in \S~\ref{summary}.

\section{Motivation, Model setup, and Simulation description}\label{basics}

In this section we describe our models of circumstellar disk with an embedded planet.
We also give a brief overview about the simulation techniques
developed in this work so that the reader can judge
the sophistication but also the kind of simplifications applied in our approach.

The primary goal of this study is to find out whether a planet or the dust in its vicinity
can be detected with observing equipment that is available now or in the near future.
Hydrodynamical simulations of gaseous, viscous protoplanetary disks with an embedded protoplanet show that the 
planet can open and maintain a significantly large gap (e.g., Bryden et al.~1999; Kley~1999, Lubow et al.~1999).
This gap, which is located along the orbit of the planet, may extend
up to a few astronomical units in width,
depending on the mass of the planet and the hydrodynamical properties of the disk.
Nevertheless, the disk mass flow on to the planet continues through
the gap with high efficiency. Nearly all of the flow through the gap is accreted by the planet at a rate comparable 
to the rate at which it would occur in the disk without the planet.

This investigation is driven by several motivations.
First, we want to investigate the possibility to observe a massive planet and the warm dust
close to it. In this respect, even a non-detection would
provide valuable constraints on planet
formation and evolution scenarios.
Second, young giant planets are expected to be very hot, compared to their old counterparts,
such as Jupiter (see, e.g., Hubbard, Burrows, \& Lunine~2002 for a recent review on the theory 
of the evolution of giant planets).
Although they are much smaller than the central star and therefore have a luminosity that is
smaller by orders of magnitude, protoplanets can still efficiently
heat the surrounding dusty environment,
as a result of the accretion or contraction process. 
Thus, the planet surroundings might be observable in 
the (far-)infrared wavelength range through thermal dust reemission.
Third, provided that the mass of the planet is large enough to open a significantly large
low-density gap (on the order of a Jupiter's mass),
the contrast between the gap and the dust heated by the planet might be sufficiently large
to distinguish both components, i.e., the gap and the dust distribution around the planet.

To achieve the above goals, we test different environments of a planet located in a circumstellar disk 
for the resulting temperature structure which,
 in combination with the density distribution,
mainly determines the likelihood to detect any of the features characterizing the embedded planet.
The models considered here cover a broad range of different, most reasonable scenarios (see \S~\ref{model}).
The following steps were performed in order to calculate the spatial temperature 
structure and density distributions:
(1) Hydrodynamical simulations of circumstellar disks with an embedded planet deriving
    the disk density structure (\S~\ref{hydro});
(2) Calculations of the large-scale temperature structure (i.e., of the whole disk) resulting from
    the stellar radiation (\S~\ref{rtsimu});
(3) Calculations of the small-scale temperature structure (in a radius of 0.5~AU around the planet)
   resulting from the additional heating by the young planet (\S~\ref{rtsimu}).
Finally, observable quantities, such as images and spectral energy distributions (SEDs),
of the whole system are derived. The single steps are described in the following sections.

\subsection{Hydrodynamics simulations}\label{hydro}

The evolution of a circumstellar disk with an embedded planet
can be formally described by the Navier-Stokes equations for the density
and the velocity field components
(see, e.g., D'Angelo, Henning, \& Kley~2002 for details). 
For the purpose of the present study, the disk is treated as a
two-dimensional viscous fluid in the equatorial, i.e., $r$--$\phi$ plane.
This means that vertically averaged quantities actually enter the 
time-dependent hydrodynamics equations. Disk material is supposed to have
a constant kinematic viscosity that is equivalent to a Shakura \& Sunyaev
parameter $\alpha=4\times10^{-3}$ at the location of the planet.
This value is consistent with those derived by calculations of
disk-planet interaction with MHD 
turbulence (Nelson \& Papaloizou~2003; Winters et al.~2003).

The model employed here is not intended to include thermal effects.
Thus, we rely on a simple equation of state:
\begin{equation}
P=c^2_s\,\Sigma,
\label{eq:pressure}
\end{equation}
where $c_s$ is the sound speed and $\Sigma$ the disk surface density. 
Note that equation~\ref{eq:pressure} yields the the vertically
integrated pressure.
As sound speed, the following approximation is used:
\begin{equation}
c_s=h\,\sqrt{\frac{G\,M_*}{r}}.
\label{eq:cs}
\end{equation}
In the above equation, $M_*$ is the stellar mass whereas $h=H/r$ indicates 
the aspect ratio of the disk, which is assumed to be constant ($H/r=0.05$).
Such value of the aspect ratio is typical for accretion disks whose 
accretion rate is on the order of 
$10^{-8}$--$10^{-7}$ ${\rm M_\sun}\,{\rm yr}^{-1}$ (e.g., Bell et
al.~1997), which is comparable to the depletion rate obtained from our numerical
simulations.

A planet-sized object with mass $M_{\rm P}$ revolves around the central star, 
moving on a circular orbit whose radius is $r_{\rm P}$. 
It perturbs the surrounding environment 
via its point-mass gravitational potential. 
The ratio of $M_{\rm P}$ to $M_*$ is $2\times10^{-3}$, hence
$M_{\rm P}= 1\,{\rm M_{\rm jup}}$ if $M_*=0.5\,{\rm M_\sun}$. 
The simulated region extends for the whole $2\,\pi$ of the azimuthal range 
and, radially, from $0.4\,r_{\rm P}$ to $4.0\,r_{\rm P}$. 
The computational domain is covered with a $242\times422$ uniform cylindrical grid.
Therefore, the resolution is such that 
$\Delta r/r_{\rm P} = \Delta \phi = 0.015$.
Because of the form in which equations are written, computation results  
can be re-scaled with respect to the parameters $M_*$, $\widehat{M}_{\rm disk}$, 
and $r_{\rm P}$, where $\widehat{M}_{\rm disk}$ represents the mass contained 
in the simulated region.  
Further details of these numerical models and related computational issues
can be found in  D'Angelo et al.~(2002).

The set of the relevant equations is solved numerically by means of a finite 
difference method which is second-order accurate in space and first-order in 
time.
The numerical algorithm is provided by an early FORTRAN-version of the code 
NIRVANA (Ziegler \& Yorke~1997) that has been improved and adapted to
the scope to perform calculations of planets in disks both in two 
and three dimensions.
We base this study on two-dimensional models because,
by means of high-resolution three-dimensional simulations,
D'Angelo, Kley, \& Henning~(2003b) 
demonstrated that computations in two dimensions give
a satisfactory description of disk-planet interactions when dealing with
heavy planets (planet-to-star mass ratio $\gtrsim 10^{-4}$) in thin 
disks ($h\approx 0.05$).

The simulations are started from a purely Keplerian disk. Due to the
angular momentum transfer among the inner disk ($r<r_{\rm P}$), the planet,
and the outer disk ($r>r_{\rm P}$), a deep density gap is soon carved in along
the orbital path. Characteristic spiral features, 
spreading both
inward and outward of the planet's orbit (see Fig.~\ref{comb6}), 
are excited by the planetary gravitational potential at locations
corresponding to Lindblad resonances. 

The disk depletion, due both to gravitational torques exerted by the planet 
on the disk material (at $r<r_{\rm P}$) and to viscous torques, is accounted
for by allowing material to drain out of the inner border of the
computational grid (neither inflow nor outflow is instead allowed through
the outer border).
Therefore, the density inside the orbit of the planet gradually reduces, 
as shown in Figure~\ref{comb6} (see also Rice et al.~2003).
This effects mimics the accretion onto the central star that,
once the system has relaxed (after a few hundred orbits), is
$3\times10^{-5}\,\widehat{M}_{\rm disk}$ per orbital period or a few times 
$10^{-9}\,{\rm M_\sun}\,{\rm yr}^{-1}$ if 
$\widehat{M}_{\rm disk}=10^{-3}\,{\rm M_\sun}$ and $r_{\rm P}$=5\,AU.

In this study we do not account for planetary migration since we deal with
massive planets orbiting in low-mass disks 
($\widehat{M}_{\rm disk}<8.5\times10^{-4}\,{\rm M_\sun}$, see Sect.~\ref{model}
for details).
When the planetary mass is larger that the local disk mass with which 
it interacts, the planet's inertia becomes important and the orbital 
evolution proceeds at a rate that is smaller than the viscous evolution 
rate of the disk (Ivanov, Papaloizou, \& Polnarev~1999). 
In fact, given the disk mass, the planet mass, and the kinematic viscosity 
employed in these calculations, the ratio between the migration timescale 
and the viscous timescale (at the planet's location) is larger than 10 and
nearly scales as $M_{\rm P}/\widehat{M}_{\rm disk}$ (hence it increases as the
disk depletes). 
Indeed, by measuring
the gravitational torques acting on the planet in our models, we find a 
migration timescale that agrees with such prediction\footnote{Although
the planet's orbit is fixed, by measuring the gravitational torques 
exerted by the disk on the planet one can evaluate the drift velocity
${\rm d}r_{\rm P}/{\rm d}t$ that the planet would have if it were allowed 
to migrate.}.
This implies that the planet's migration would occur on a timescale longer 
than $10^6$ years, which is comparable with the timescale of disk dispersal.

\subsubsection{Planetary accretion luminosity}\label{sec:PAL}

These calculations also simulate the growth rate of the planet due to
the feeding process by its surroundings.
This is achieved according to the procedure outlined in D'Angelo et al.~(2002). 
A fraction of the matter orbiting the planet inside of an accretion region is removed.
Since the accretion process is a highly localized phenomenon,
the accretion region must be small. Therefore, we choose a radius 
$r_{\rm P}^{\rm acc}$ equal to 
$6\times10^{-2}\,r_{\rm H}$, where the Hill radius
$r_{\rm H}=r_{\rm P}\,\sqrt[3]{M_{\rm P}/(3\,M_*)}=8.7\times10^{-2}\,r_{\rm P}$
approximately characterizes the sphere of gravitational influence of the 
planet\footnote{While larger accretion regions would only be justified
by poor numerical resolutions, computations performed with shorter
lengths of $r_{\rm P}^{\rm acc}$ and appropriate resolution furnish
accretion rates comparable to the one obtained in these calculations.}.
In order to achieve the necessary resolution to study the flow dynamics on 
these short length scales, without neglecting the global circulation within 
the disk, a nested-grid technique has been utilized. 
This numerical strategy permits to capture, at the same time, 
large as well as small length and timescales of the problem. 
Therefore, it allows to obtain an accurate 
evaluation of the planetary accretion rate $\dot{M}_{\rm P}$. 
In fact, the portions of the disk nearest to the planet are resolved with a 
mesh step equal to 
$\Delta r_{\rm min}/r_{\rm P} = \Delta \phi_{\rm min}=1.8\times10^{-3}$.
After the hydrodynamical variables (density and velocities) have reached a 
quasi-stationary state (300--400 orbital periods after the beginning of the
simulation), for $M_{\rm P}/M_* = 2\times10^{-3}$
the measured value of $\dot{M}_{\rm P}$ is $\simeq 2\times10^{-5}$
$\widehat{M}_{\rm disk}$ per orbit, 
where $\widehat{M}_{\rm disk}$ represents the disk mass comprised in the
computational domain.
For an object at 5 AU and a disk mass within $20$ AU of 
$8.5\times10^{-4}\,{\rm M_\sun}$ 
(see \S~\ref{model}), one gets $\dot{M}_{\rm P}\approx 1.5\times10^{-6}$
${\rm M_{\rm jup}}\,{\rm yr}^{-1}$, which is evaluated at an evolutionary time 
$\approx4\times10^{3}$ years. 
As expected, this value is smaller than that obtained for a Jupiter-mass planet
orbiting a solar-mass star (see D'Angelo et al.~2003b) because the density 
gap along the orbit is wider and deeper.

Since a measure of the planetary accretion rate is available, it is possible 
to evaluate the accretion luminosity of the planet $L^{\rm acc}$.
For such purpose we assume that the planet has a gravitational potential
of the form $\Phi_{\rm P}=-G\,M_{\rm P}/s$, where $s$ is the distance from
the planet. Indicating with
$S_{\rm P}$ the planet's radius, which is on the order of a few Jupiter radii
(e.g., Burrows et al.~1997), we can then write
\begin{equation}
L^{\rm acc}=-\dot{M}_{\rm P}\left[%
           \Phi_{\rm P}(S_{\rm P})-\Phi_{\rm P}(r_{\rm P}^{\rm acc})\right],
\end{equation}
which becomes
\begin{equation}
L^{\rm acc}= G\,M_{\rm P}\,\dot{M}_{\rm P}\left(%
           \frac{1}{S_{\rm P}}-\frac{1}{r_{\rm P}^{\rm acc}}\right).
\label{eq:lacc}
\end{equation}
If we replace all the constants and variables in equation~\ref{eq:lacc} 
with
the appropriate numbers, for a planet with $M_{\rm P}= 1\,{\rm M_{\rm jup}}$ 
(orbiting a $M_{\rm *} = 0.5\,{\rm M_{\sun}}$ star, see \S~\ref{model})
and $r_{\rm P}=5$\,AU, we get 
$L^{\rm acc} \approx 10^{-4}\,{\rm L_{\sun}}$.
However, the accretion rate is observed to slowly diminish with the 
time, due to the depletion of the disk and the deepening of the gap. 
Hence, the accretion luminosity decays with the time. 
As an example, extrapolating $\dot{M}_{\rm P}$ after a few $10^4$
years, the above estimate of $L^{\rm acc}$ becomes 
$\approx 10^{-5}\, {\rm L_{\sun}}$.
This value for the planetary accretion luminosity is on the same order of
magnitude as the one derived by Burrows et al.(1997) for a young Jupiter-mass
planet.

\subsection{Radiative transfer}\label{rtsimu}

\begin{figure}[t]
  \epsscale{0.4}
  \plotone{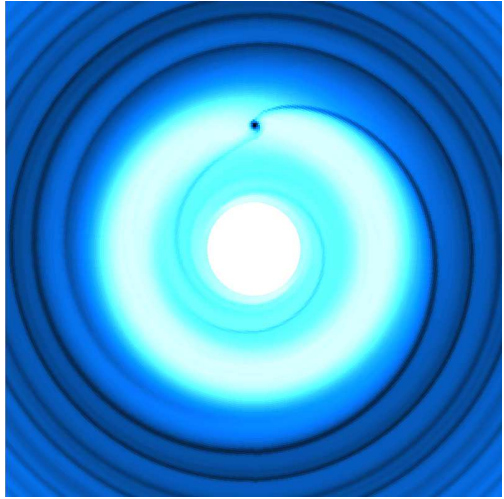}
  \caption{Midplane density distribution of the inner region of a disk.
    The side length of the image is 4\,$r_{\rm P}$, and thus
    4\,AU/20\,AU in case of a planet orbiting the star at a distance
    of 1\,AU/5\,AU. The image refers to a dynamical time of $350$ orbits. 
    High/low density regions are represented by a dark/bright colors.
    Note the spiral density waves caused by the planet's gravitational
    potential and the low density
    region inside the planet's orbit caused by the transfer of angular
    momentum (between the disk and the planet) and the ongoing accretion onto the central star.}
  \label{comb6}
\end{figure}

Based on the density structure obtained in the hydrodynamical simulations described above
we derive a self-consistent temperature structure in the disk.
For this purpose we use the three-dimensional continuum radiative transfer code MC3D
(Wolf~2003; see also Wolf et al.~1999).
The spatial discretization of the density and temperature in the disk midplane was
chosen to be identical to the grid used in the hydrodynamical simulations.
In addition, the density structure in vertical direction has a Gaussian profile
\begin{equation}
\rho(r,\phi,z) \propto
\Sigma(r,\phi) 
\exp\left[-\left(\frac{z}{H(r,\phi)}\right)^2\right],
\label{eq:rho}
\end{equation}
where
$\Sigma(r,\phi)$ is the surface density resulting from the
hydrodynamical calculations and
$H(r,\phi)$ is the scale height above the midplane. 
A realistic shape of $H$ would present a trough along the gap because
of the low temperatures and the gravitational attraction of the
planet (D'Angelo et al. 2003a). Therefore, matter around the planet is partly
shielded from direct stellar irradiation, if the density at high latitudes
inside of the planet's orbit is high enough.
It is worthwhile to stress here that, although we use a constant
aspect ratio $H/r$, the mass density $\rho$ does have a trough at $r=r_{\rm P}$
since the surface density $\Sigma$ (the leading term in
equation~\ref{eq:rho}) is lower in the gap region that anywhere else in the
disk (see Fig.~\ref{comb6}). 
Thus, even though $H/r$ is a constant, the shielding effect is accounted for.

In order to achieve a similar resolution of the temperature structure both in vertical and radial direction 
we use a spherical grid $(r,\theta,\phi)$ with an opening angle of $\approx 6^{\rm o}$.
Using a discretization of 35 grid points in $\theta$-direction, the density/temperature
grid consists of $\approx1.8\times10^6$ cells. This grid is used to derive the temperature structure
in the disk as resulting from stellar radiation. The luminosity, resulting
from possible accretion onto the star is not accounted for separately.

In order to consider the additional contribution of the planet to the heating of its local environment, 
the resolution has to be increased locally, within a volume which can be efficiently heated
by the planet. The combination of the grid covering the whole disk and a second grid allowing 
for the simulation of a highly resolved temperature structure in the vicinity of the planet is
numerically possible within the applied radiative transfer scheme. However, this approach
would require a huge amount of computing time in order to derive the temperature structure
on the second grid resulting from stellar heating. Furthermore, we place the planet
at a distance of either 1\,AU or 5\,AU from the star (see \S~\ref{model}). Thus, the temperature gradient
at the position of the planet is mainly determined by the planet's contribution, while
the stellar heating only results in minor temperature variations in the circumplanetary region.
This allows us to decouple the heating by the star from that by the planet. Therefore,
we derive the stellar temperature structure on the ``low-resolution'' grid as described above
covering the whole disk, but we simulate the additional heating due to
the planet on a much smaller grid centered 
on the planet, which sufficiently resolves the temperature gradient in its vicinity.

The grid centered on the planet is spherical. The local temperature structure is calculated in three dimensions:
$T_{\rm P}(r_{\rm P},\theta_{\rm P},\phi_{\rm P})$ whereby the grid consists of 
$(n_{\rm r, P}=100) \times (n_{\rm \theta, P}=35) \times (n_{\rm \phi, P}=72)$ grid points. 
It has a linearly equidistant step width in $\theta$ and $\phi$ direction,
while the discretization in radial direction is defined by
$(r_{i+2}-r_{i+1})/(r_{i+1}-r_{i})=1.1$ $(i=1,2,...,n_{\rm r,P})$, where $r_{i}$ are 
the radial coordinates of the grid cell boundaries. The inner radius of the grid is the radius
of the planet and therefore depends on the particular model (see \S~\ref{model}). The outer radius 
is fixed to 0.5~AU, chosen according to the maximum simulated effective temperature and luminosity of the planet:
$T_{\rm P}$=1790\,K and $L_{\rm P} = 10^{-3.4}{\rm L_{\sun}}$, respectively.
The contribution of the planet to the heating of the material outside this region is negligible
compared to the heating by the star. For reasons of simplification, the SED of the planet is considered 
to be that of a blackbody with the temperature $T_{\rm P}$.

Dust grains are the dominant source of absorption and emission in the circumstellar disk.
The dust grains are assumed spherical, consisting of a mixture of 62.5~\% astronomical silicate
and 37.5~\% graphite 
(optical data from Weingartner \& Draine~2001)\footnote{For graphite 
we adopt the usual ``$\frac{1}{3}-\frac{2}{3}$'' approximation 
(Draine \& Malhotra~1993):
$Q_{\rm ext} = [Q_{\rm ext}(\epsilon_{\parallel}) + 2 Q_{\rm ext}(\epsilon_{\perp})]/3$,
where $\epsilon_{\parallel}$ and $\epsilon_{\perp}$ are the components of the graphite
dielectric tensor for the electric field parallel and perpendicular to the crystallographic $c$-axis,
respectively ($Q_{\rm ext}$ is the extinction efficiency factor).}.
The size distribution $n(a)$ of the grains follows a power law, $n(a) \propto a^{-3.5}$, with
grain radii in the range $0.005\,\mu{\rm m} \le a \le 1\,\mu{\rm m}$.
The gas-to-dust mass ratio amounts to 100:1 in our simulations.

\subsection{Model setup}\label{model}

We now define a set of disk models that covers a large range of possible configurations.
Each model is based on the density distribution of a circumstellar
disk containing a planet, as described in \S~\ref{hydro}.
This approach is possible because the results of our hydrodynamical simulations are scale-free, 
as mentioned above. Thus, one can change the total mass and size of
the disk through the parameters $\widehat{M}_{\rm disk}$ and 
$r_{\rm P}$ (as long as the assumption of negligible
self-gravity is valid) whereas the mass of the planet can be changed
via the parameter $M_*$ (since the planet-to-star mass ratio is fixed).

We investigate the following parameter space of disk/planet configurations:
\begin{enumerate}
\item   {\bf Masses of the planet and the star:}
  $M_{\rm P} = 1\,{\rm M_{\rm jup}}$ and $5\,{\rm M_{\rm jup}}$.
  The corresponding mass of the star amounts to $M_{\rm *} = 0.5\,{\rm M_{\sun}}$ and $2.5\,{\rm M_{\sun}}$,
  respectively (in order to represent a T\,Tauri type and a Herbig Ae type star).

\item   {\bf Distance of the planet from the star:}
  $r_{\rm P}$ = 1\,AU and 5\,AU.

\item   {\bf Mass of the disk:}
        We consider disks with masses of $M_{\rm disk} = 10^{-2}, 10^{-4}$, and $10^{-6}\,M_{\sun}$.
        The first two masses cover typical masses of dust expected and found in T~Tauri and Herbig Ae/Be
        disks (Shu et al.~1987, Beckwith et al.~1990; 
        see also McCaughrean et al.~2000, Mundy et al.~2000, Natta et al.~2000,
        Wilner \& Lay~2000 and references therein). A much lower mass of $10^{-6}\,M_{\sun}$, however,
        is considered in our investigation as well in order to account for more evolved disks.

        Depending on the planet distances $r_{\rm P}$ = 1\,AU and $r_{\rm P}$ =5\,AU, our disk models
        have a radius of 4\,AU and 20\,AU, respectively. However,
        this  represents the inner disk region, 
        containing only a fraction of the masses of the dust and gas given above. We base our assumptions about the
        mass contained in this inner region on the model of the circumstellar disk of the T\,Tauri star
        IRAS~04302+2247 (Wolf et al.~2003). Within a radius of 4\,AU
        (or 20\,AU) this model contains 
        1.9\% (or 8.5\%) of the total disk mass, $M_{\rm disk}$ 
        (i.e., $\widehat{M}_{\rm disk}<8.5\times 10^{-4}\,{\rm M_\sun}$).

\item   {\bf Planetary and Stellar parameters:}
        In order to account for the additional heating of the dust in
        the immediate vicinity of the planet,
        we use luminosities (and temperatures) of the youngest planets in the evolutionary scenario
        described by Burrows et al.~(1997):
        $10^{-4.8} L_{\sun}$ / 830 K for the $M_{\rm P} =1\,{\rm M_{\rm jup}}$ planet and
        $10^{-3.4} L_{\sun}$ / 1790 K for the $M_{\rm P} =5\,{\rm M_{\rm jup}}$ planet.
        These values are also in rough agreement (or at least not smaller than) the accretion luminosities
        approximately derived in Sect.~\ref{sec:PAL}, i.e., we consider the most optimistic scenario
        for the planetary heating.
        For the stellar parameters we use typical luminosities / temperatures for 
        a T\,Tauri star ($L_{\rm *} =  0.92\,{\rm L_{\sun}}$, T=4000\,K for $M_{\rm *} = 0.5\,{\rm M_{\sun}}$)
        and
        a Herbig Ae star ($L_{\rm *} =  46\,{\rm L_{\sun}}$, T=9500\,K for $M_{\rm *} = 2.5\,{\rm M_{\sun}}$),
        respectively.
\end{enumerate}

\section{Results}\label{results}

We now discuss observable quantities that we derive from the density / temperature structure for each configuration
of the circumstellar disk. We assume the disk to be seen face-on since this orientation allows
a direct view on the planetary region.

The thermal dust reemission of circumstellar disks can be mapped with (sub)millimeter observations
and it has been performed successfully for a large number of young stellar objects 
(e.g., Beckwith et al.~1990).
The only observatory, however, which - because of its aspired resolution and sensitivity -
might have the capability in the near future to detect features as small as a gap
induced by a planet in a young disk, will be the Atacama Large Millimeter 
Array\footnote{See \texttt{http://www.arcetri.astro.it/science/ALMA/} for a compilation of documents related to ALMA.}.
For this reason, we perform radiative transfer simulations in order to obtain images at a frequency of 900\,GHz,
which  marks the planned upper limit of the frequency range to be covered by ALMA.
This frequency is required for our simulations because it allows to
obtain the highest spatial resolution.
The trade-off is the relatively high system temperature ($\sim$ 1200\,K; Guilloteau~2002)
which adds a significant amount of noise to the simulated observations.
Therefore, long observing times (8h, in our simulations) are mandatory.
To investigate the resulting maps to be obtained with ALMA, 
we use the 'ALMA simulator' developed by Pety, Gueth, \& Guilloteau~(2001)
and chose an observation setup according to the directions and suggestions given by Guilloteau~(2002).
Furthermore, we introduce 
a random pointing error during the observation with a maximum value of 0.6'' in each direction,
$30^{\rm o}$ phase noise, and further error sources, such as amplitude errors and 
``anomalous'' refraction (due to the variation of the refractive index of the wet air along
the line of sight).
The observations are simulated for continuous observations centered on the meridian transit
of the object. The object passes the meridian in zenith where an opacity of 0.15 is assumed.
The bandwidth amounts to 8\,GHz.

Based on these simulations, whose outcomes are shown in Figures~\ref{alma1}-\ref{seds}, 
we make the following predictions about the observability of a giant
planet in a young circumstellar disk:
\begin{enumerate}
\begin{figure}[t]
  \epsscale{1.0}
  \plotone{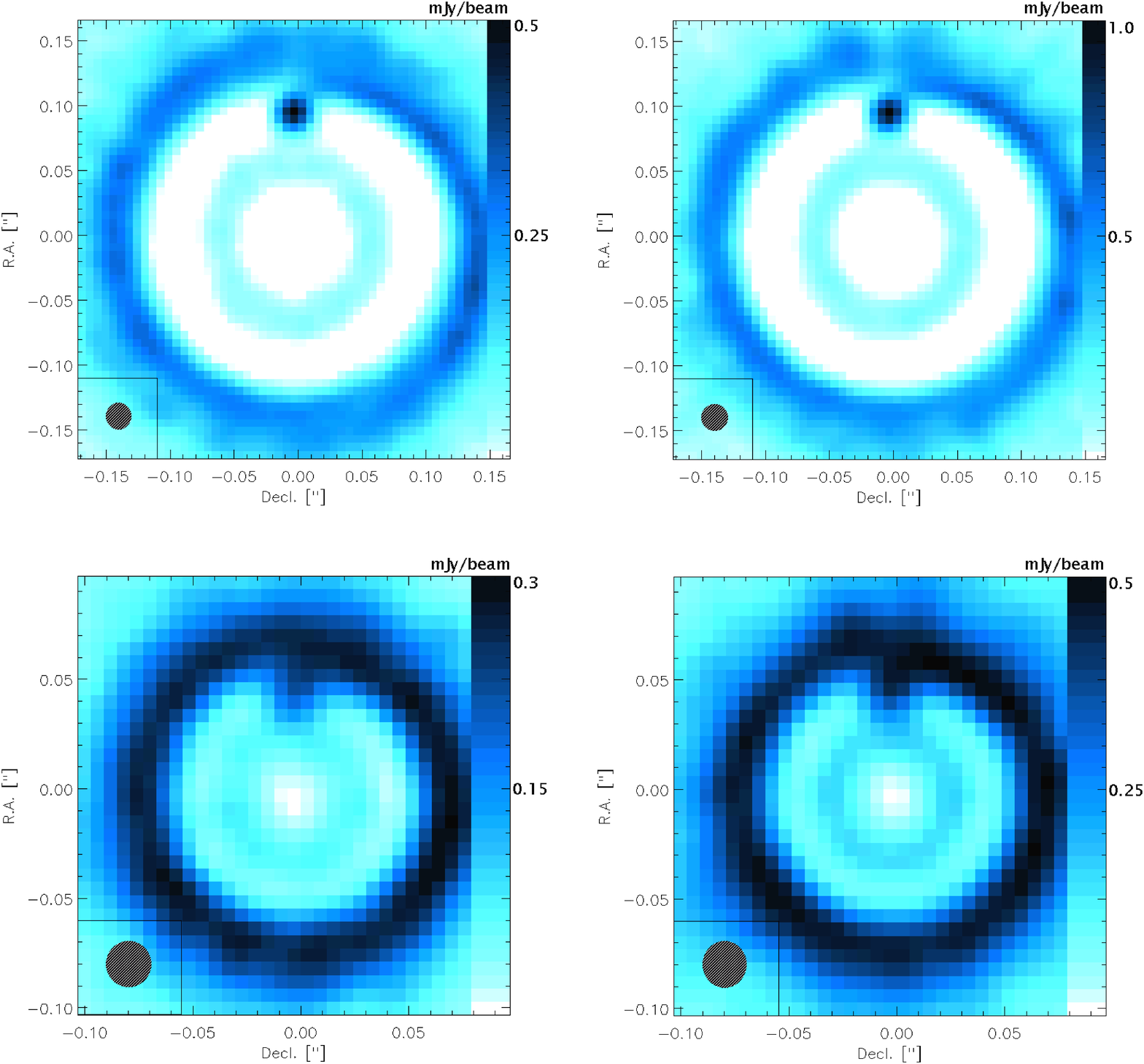}
  \caption{
    Simulation of ALMA observations of disk with 
    an embedded planet of $1\,{\rm M_{\rm jup}}$ (left column) / $5\,{\rm M_{\rm jup}}$ (right column)
    around a  $0.5\,{\rm M_{\sun}}$ / $2.5\,{\rm M_{\sun}}$ star 
    (orbital radius: 5\,AU).
    The assumed distance is 50\,pc (top row) / 100\,pc (bottom row).
    The disk mass amounts to $M_{\rm disk} = 1.0\times10^{-2}\,{\rm M_{\sun}}$.
    Only structures above the $2\sigma$-level are shown.
    The size of the combined beam is symbolized in the lower left edge of each image.
    Note the reproduced shape of the spiral wave near the planet and the slightly shadowed region
    behind the planet in the upper images.
  }
    \label{alma1}
\end{figure}
\item The resolution of the images to be obtained with ALMA
will allow detection of the warm dust in the vicinity of the planet only if the object is at a distance
of not more than about 50-100\,pc (see Fig.~\ref{alma1}). 
For larger distances, the contrast between the planetary
region and the adjacent disk in any of the considered planet/star/disk configurations will be too low to be detectable.
\item Even at a distance of 50\,pc a resolution being high enough to allow a study of the circumplanetary region 
can be obtained only for those configurations with the planet on a Jupiter-like orbit but not when it 
is as close as 1\,AU to the central star. This is mainly due to the size of the beam 
($\sim 0.02''$ for ALMA setup as described above).
%
%
\begin{figure}[t]
  \epsscale{1.0}
  \plotone{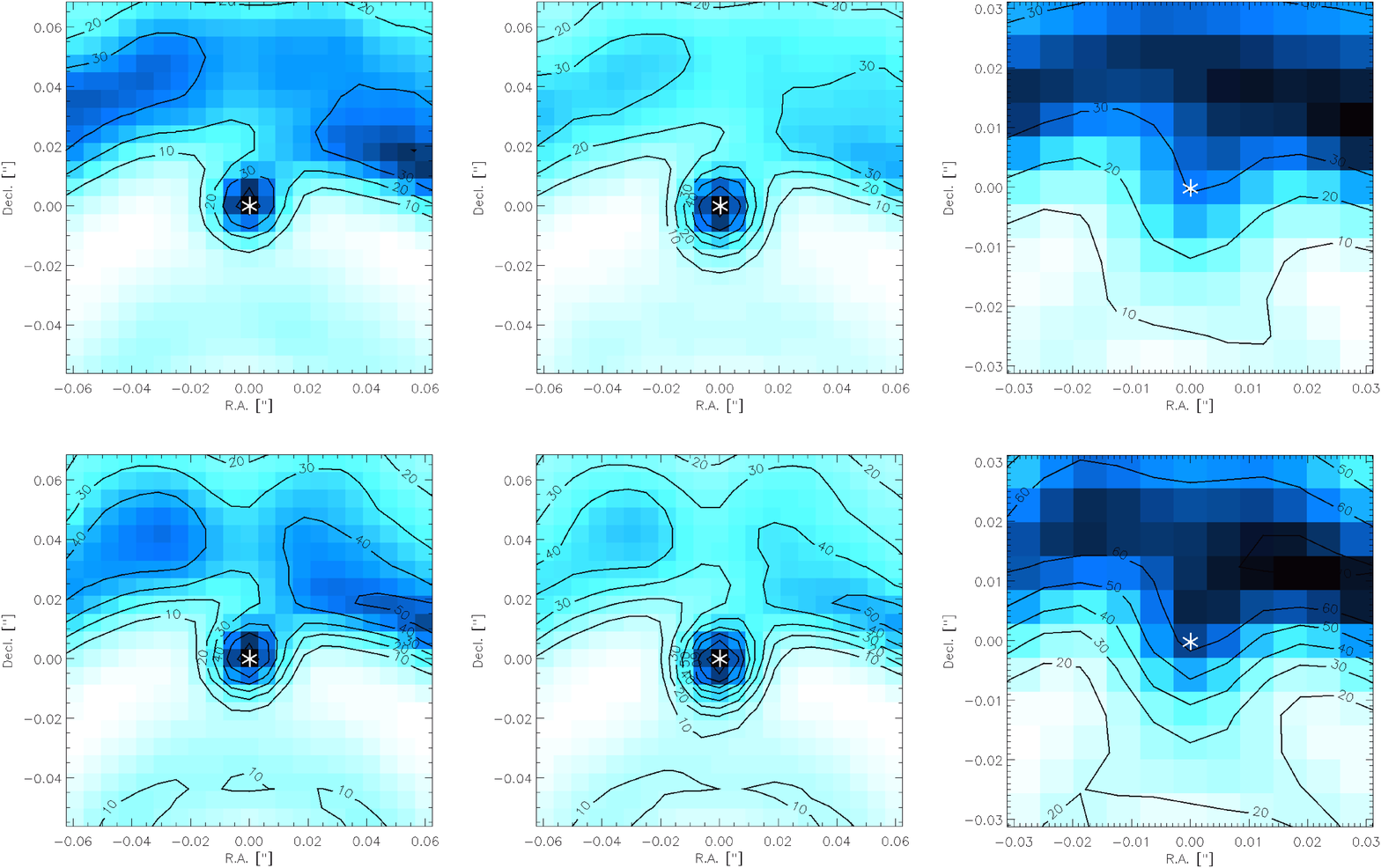}
  \caption{
    Simulation of ALMA observations with a close-up view of the region around the planet.
    Top row:    $M_{\rm P} / M_{\rm *} = 1\,{\rm M_{\rm jup}} / 0.5\,{\rm M_{\sun}}$,
    bottom row: $M_{\rm P} / M_{\rm *} = 5\,{\rm M_{\rm jup}} / 2.5\,{\rm M_{\sun}}$.
    The left and middle column: 
    Comparison of the case without (left column) and with (right column) additional heating
    in the planetary region; the distance of the disk is 50\,pc 
    (see \S~\ref{results} for a more detailed description).
    Middle and left column:
    Comparison of disks, seen at two different distances (middle column: 50\,pc, right column: 100\,pc).    
    The size of the displayed region amounts to 6\,AU$\times$6\,AU.
    The asterisk marks the position of the planet.}
  \label{alma2}
\end{figure}

\item Both heating sources described in \S~\ref{rtsimu} have been found to be of importance for warming up
the dust in the vicinity of the planet so that it can be detectable.
The influence of the planetary radiation is demonstrated in Figure~\ref{alma2}, showing 
the case of a highly-luminous planet in comparison with the case of no significant planetary heating.
The importance of the second heating source, i.e., the radiation from the central star, increases 
(relative to the planetary radiation),
with increasing mass (luminosity) of the star if - as in our simulations - the stellar to planetary mass
ratio is considered to be constant.
This can be explained by the values of the stellar and planetary luminosity given in \S~\ref{model}.
The luminosity ratio $L_{\rm *} / L_{\rm P}$ increases by about a factor of 2 by increasing
the mass of the star/planet from $0.5\,{\rm M_{\sun}} / 1\,{\rm M_{\rm jup}}$ 
to $2.5\,{\rm M_{\sun}} / 5\,{\rm M_{\rm jup}}$.
The evolution of the disk within the planet's orbit is also very important
for the relative contribution of the stellar heating. 
While we consider the inner disk to be significantly removed in our simulations, allowing
the heating of the planetary environment also by the star, less evolved disks will efficiently
shield the planet from the stellar radiation.
%
\item The observation of the emission from the dust in the vicinity of the planet will be possible only
in case of the most massive circumstellar disks.
While the signal of the planet has a strength of $\sigma$=89.7/65.3 (with respect to the background noise,
not the average signal of the disk!) in case of $M_{\rm disk} = 10^{-2}\,M_{\sun}$ and a planet 
mass $1\,{\rm M_{\rm jup}}$/$5\,{\rm M_{\rm jup}}$,
it is down to only $\sigma$=1.1/0.4 for the corresponding configuration
with $M_{\rm disk} = 10^{-4}\,M_{\sun}$.
\end{enumerate}

\begin{figure}[t]
  \epsscale{0.5}
  \plotone{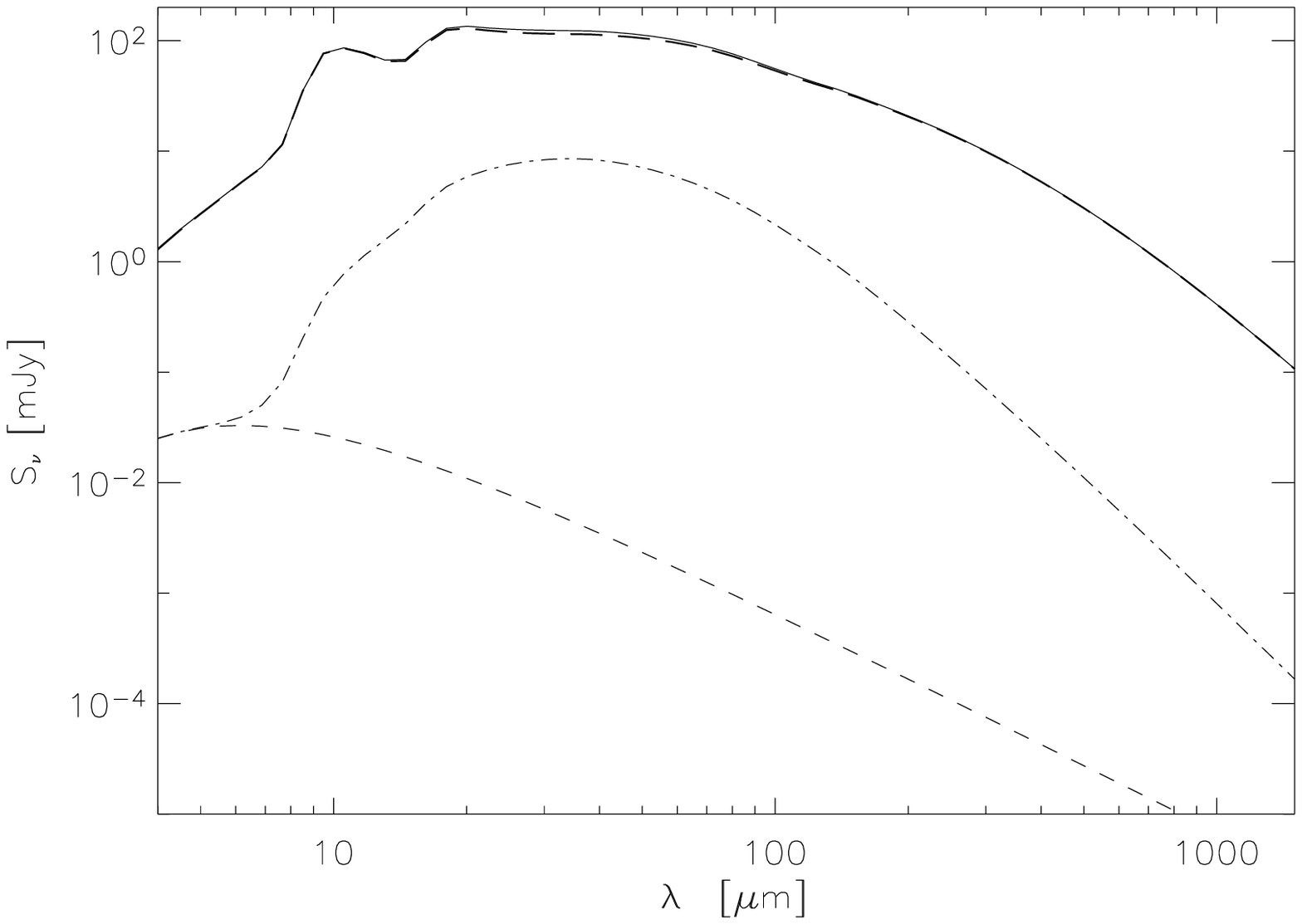}\\
  \plotone{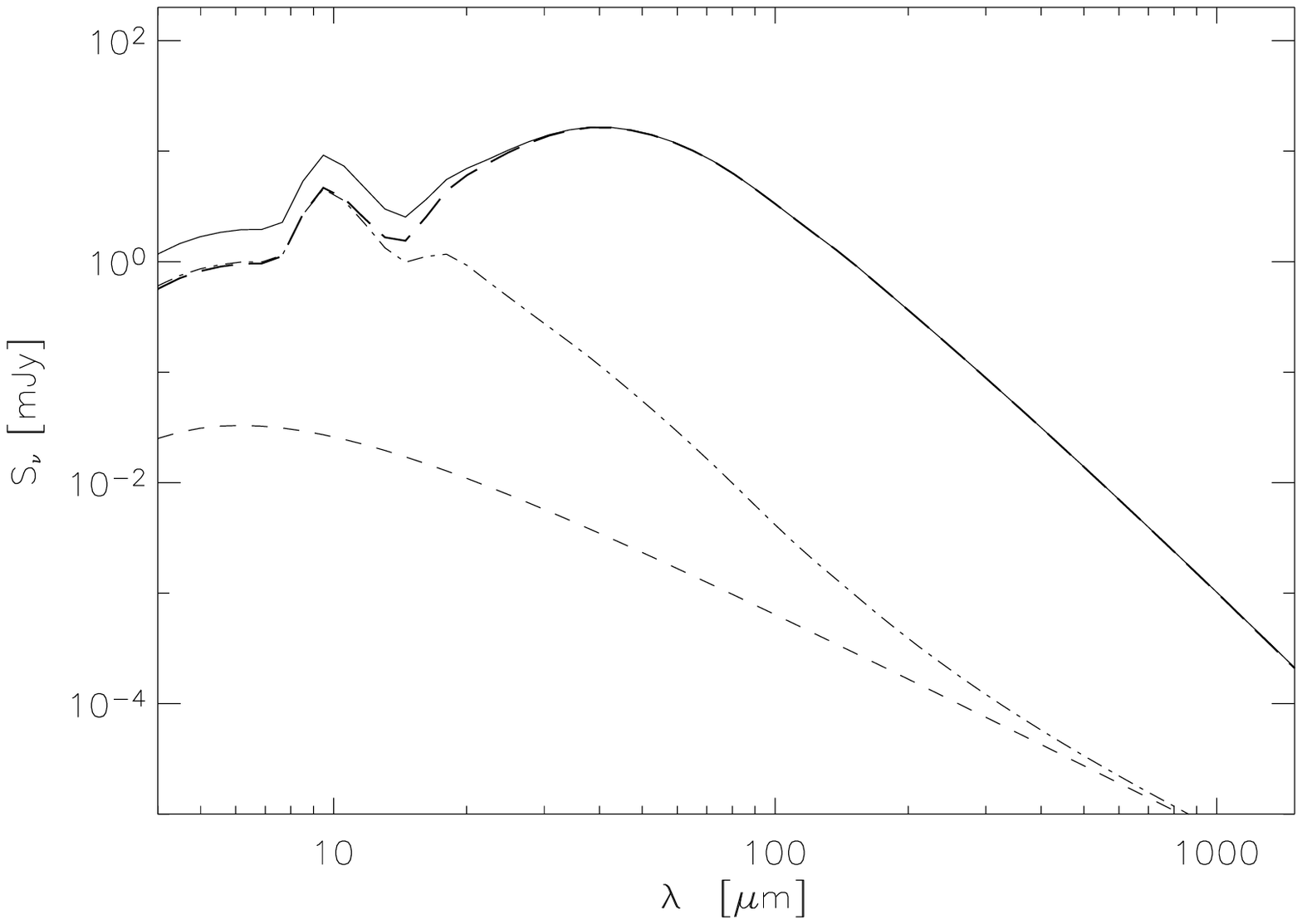}\\
  \plotone{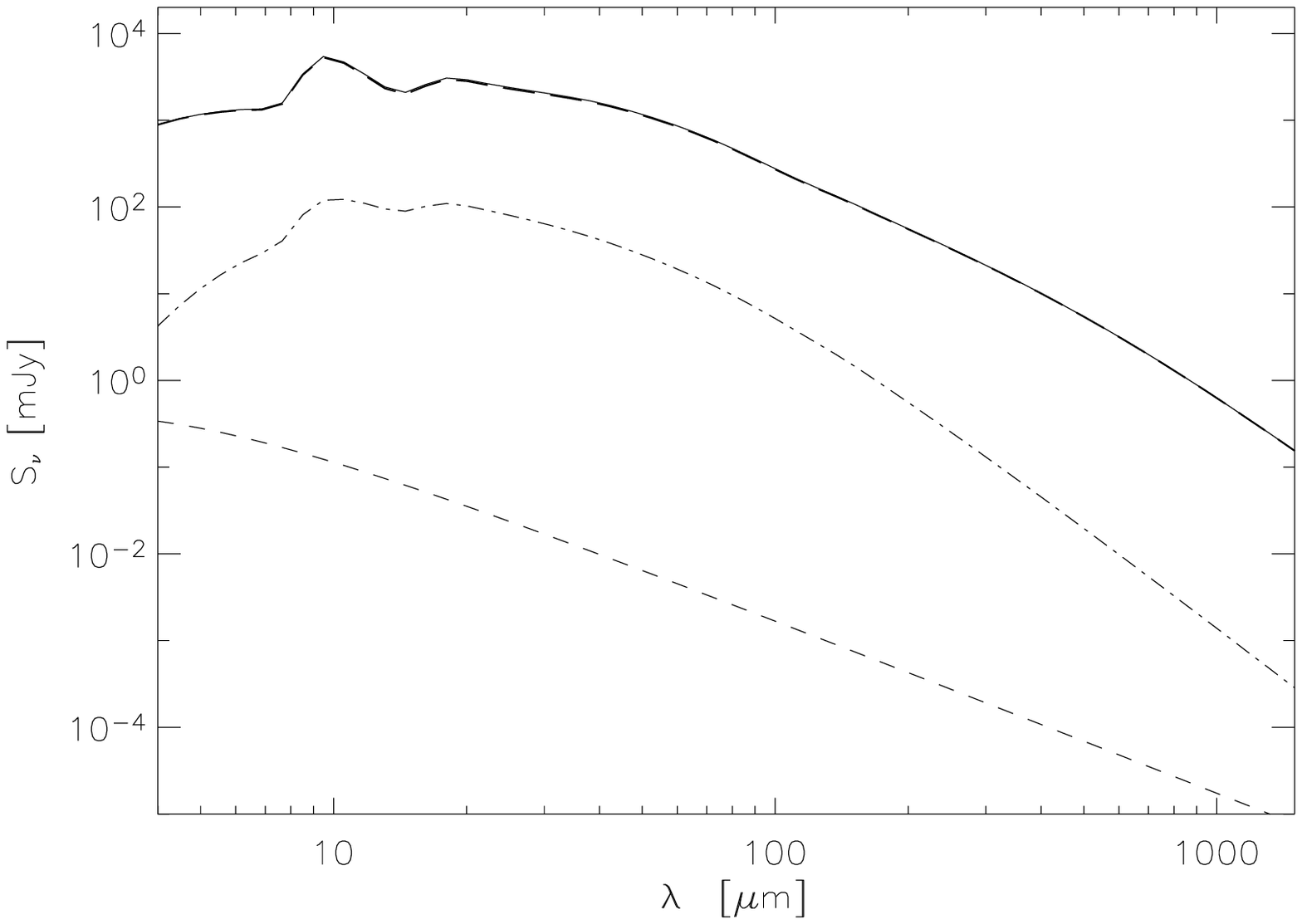}
  \caption{
  Spectral energy distribution of selected disk models.
{\em Top:} 
  $M_{\rm P} / M_{\rm *} = 1\,{\rm M_{\rm jup}} / 0.5\,{\rm M_{\sun}}$,
  $M_{\rm disk} = 10^{-2}\,M_{\sun}$ 
  (same model as used for Fig.~\ref{alma2}, top row / middle column);
{\em Middle:} 
  $M_{\rm P} / M_{\rm *} = 1\,{\rm M_{\rm jup}} / 0.5\,{\rm M_{\sun}}$,
  $M_{\rm disk} = 10^{-6}\,M_{\sun}$;
{\em Bottom:} 
  $M_{\rm P} / M_{\rm *} = 5\,{\rm M_{\rm jup}} / 2.5\,{\rm M_{\sun}}$,
  $M_{\rm disk} = 10^{-2}\,M_{\sun}$.
  The planet is located at a distance of 5\,AU from the central star.
  {\em Solid line}:        net SED;
  {\em long dashed line}:  disk reemission;
  {\em short dashed line}: direct (attenuated) and scattered radiation
  originating from the protoplanet (due to planetary radiation and/or accretion);
  {\em dot-dashed line}:   (re)emission from the planet and the dust within the sphere 
                           with a radius of 0.5\,AU centered on the planet.
  The assumed distance is 140\,pc.
  For comparison, the photospheric fluxes at 10/100/1000$\mu$m are
  30/0.35/0.0036\,mJy (125/1.3/0.013\,mJy) in case
  of the low-mass (high-mass) star.
  }
  \label{seds}
\end{figure}
We want to remark that the contribution from the planet to the net flux at 900\,GHz
-- by direct or scattered radiation and reemission of the heated dust in its vicinity --
is $\le 0.4\%$ (depending on the particular model) than that from the small region of the disk 
considered in our simulations.
Furthermore, the planetary radiation significantly affects the dust reemission SED only in the 
near to mid-infrared wavelength range (see Fig.~\ref{seds}, upper two SEDs). 
However, since this spectral region is influenced also by the warm 
upper layers of the disk and the inner disk structure, the planetary contribution 
and thus the temperature / luminosity of the planet cannot be derived from the SED.
In case of the massive planet / star ($M_{\rm P} / M_{\rm *} = 5\,{\rm M_{\rm jup}} / 2.5\,{\rm M_{\sun}}$),
the influence of the planet is even less pronounced in the mid-infrared wavelength range 
(see Fig.~\ref{seds}, lower SED), which can be explained by the lower luminosity 
ratio $L_{\rm P} / L_{\rm *}$, as discussed above.

\section{Conclusions}\label{summary}

We performed simulations with the goal to answer the question of whether the circumplanetary
environment of a giant planet, embedded in a young circumstellar disk, can be detected.
Indeed, we could show that a detection will be possible 
via mapping with the Atacama Large Millimeter Array, even if under particular circumstances.
We found two major constraints that limit the applicability of this approach to study planet formation.
First, only for nearby objects at distances of typically less than $\approx 100$\,pc the spatial resolution
provided by ALMA will be high enough to allow to spatially separate the circumstellar disk and circumplanetary
material from one another. Second, the emission from the circumplanetary environment is significantly strong only in the case
of massive and thus young circumstellar disks.
Furthermore, the signal of the radiation from the planet / circumplanetary material will be best distinguishable
if a high planet-to-star mass ratio is targeted.
We want to remark that in contrast to the most successful planet detection technique so far,
based on radial velocity measurements (e.g., Mayor \& Queloz~1995, Marcy \& Butler~2000),
the likelihood of detection does not increase with decreasing distance of the planet to the star.
In particular, in none of the model configurations with a planet at a
distance of only 1\,AU from the central
star (as opposed to 5\,AU in models with positive detections) 
the circumplanetary environment could be distinguished from the circumstellar disk.

As our study shows, high-resolution (sub)millimeter observations will provide a valuable tool in the near future
to challenge planet formation theories. For this reason it is worth to continue this investigation 
with the goal to achieve more detailed predictions.
There are three major issues which could be significantly improved:
First, the simulation of the radiative transfer (or the energy transfer in general) should be coupled
with or embedded in the hydrodynamical calculations.
According to the existing computational resources this goal can be achieved in the near future 
(see, e.g., D'Angelo et al.~2003a).
Second, the hydrodynamical simulations that allow to gain a higher resolution in the region of the planet
should be employed (e.g., D'Angelo et al.~2003b; Bate et al.~2003).
In this manner the dust reemission could be calculated more accurately.
Third, the emission spectrum of the protoplanet should be modeled more accurately:
Internal heating and heating due to accretion onto the planet have to be investigated separately
(taking into account the anisotropy of different radiative sources).
Furthermore, objects with temperatures $\sim 10^3$\,K are expected to have strong 
absorption bands from molecular transitions and even dust absorption solid state features
in their spectra (e.g., Allard et al.~2001), which should be considered in the heating process 
of the local planetary environment.

\acknowledgments

S.~Wolf was supported through the DFG Emmy Noether grant WO 857/2-1, the NASA grant NAG5-11645 and through 
the SIRTF/Spitzer legacy science program through an award issued by JPL/CIT under NASA contract 1407.
S.~Wolf thanks G.~Bryden for helpful discussions, F.~Gueth for support during the work with the ALMA simulator,
J$\acute{\rm e}$r$\hat{\rm o}$me Pety and Katharina Schreyer for their help with GILDAS,
and A.~Hatzes for his support in obtained computing time at Computing Center of the Friedrich Schiller University 
(Jena, Germany), where some of the simulations have been performed.
The hydrodynamical calculations reported here were performed using the
UK Astrophysical Fluids Facility (UKAFF).

\appendix


\end{document}